\documentclass{article}
\usepackage[utf8]{inputenc}
\usepackage{graphicx}
\usepackage{amssymb}
\usepackage{amsthm}
\usepackage{amsmath}
\usepackage{hyperref}
\usepackage{color}
\usepackage{microtype}
\usepackage{authblk}

\newcommand{\bigoh}{\mathcal{O}}
\newcommand{\littleoh}{o}
\newcommand{\app}{$\star$}
\newcommand{\compl}[1]{\overline{#1}}

\newenvironment{qedproof}{\begin{proof}}{\qed\end{proof}}

\DeclareMathOperator{\better}{\preceq}
\DeclareMathOperator{\conc}{\oplus}
\DeclareMathOperator{\cert}{cert}
\DeclareMathOperator{\sm}{sm}
\DeclareMathOperator{\mm}{mm}
\DeclareMathOperator{\smw}{smw}
\DeclareMathOperator{\mmw}{mmw}
\DeclareMathOperator{\act}{act}
\DeclareMathOperator{\tot}{tot}

\newtheorem{theorem}{Theorem}
\newtheorem{lemma}[theorem]{Lemma}

\newtheorem{observation}[theorem]{Observation}
\newtheorem{corollary}[theorem]{Corollary}

\newtheorem{proposition}[theorem]{Proposition}

\title{Solving Hamiltonian Cycle by an EPT Algorithm for a Non-sparse
Parameter
}
\author{Sigve Hortemo S\ae{}ther\thanks{Supported by the Norwegian Research Council.}}
\affil{University of Bergen}
\date{}

\begin{document}

\maketitle

\begin{abstract}
  Many hard graph problems, such as Hamiltonian Cycle, become FPT when
  parameterized by treewidth, a parameter that is bounded only on sparse graphs.
  When parameterized by the more general parameter clique-width, Hamiltonian Cycle
  becomes W[1]-hard, as shown by Fomin et al.~\cite{W1hardness}. S\ae{}ther and
  Telle address this problem in their paper~\cite{ST} by introducing a
  new parameter, split-matching-width, which lies between treewidth
  and clique-width in terms of generality. They show that even though
  graphs of restricted split-matching-width might be dense, solving
  problems such as Hamiltonian Cycle can be done in FPT time.

  Recently, it was shown that Hamiltonian Cycle parameterized by
  treewidth is in EPT~\cite{jesper,matroidReps}, meaning it can be
  solved in $n^{\bigoh(1)}2^{\bigoh(k)}$-time. In this paper, using
  tools from~\cite{matroidReps}, we show that also parameterized by
  split-matching-width Hamiltonian Cycle is EPT. 
  To the best of our knowledge, this is the first EPT algorithm for any
  "globally constrained" graph problem parameterized by a non-trivial and
  non-sparse structural parameter.   To accomplish this,
  we also give an algorithm constructing a branch decomposition
  approximating the minimum split-matching-width to within a constant
  factor. Combined, these results show that the algorithms in
  \cite{ST} for Edge Dominating Set, Chromatic Number and Max Cut all
  can be improved. We also show that for Hamiltonian Cycle and Max Cut the resulting
  algorithms are asymptotically optimal under the 
  Exponential Time Hypothesis.

\end{abstract}

\section{Introduction}

The problem of finding a Hamiltonian Cycle in a graph - a simple cycle
covering all the vertices of the graph - is NP-complete~\cite{gareyJohnson}.
One way to handle an NP-hard problem is by investigating its parameterized
complexity, for various choices of parameter.
Unlike a lot of other NP-hard graph problems, Hamiltonian Cycle does not
have a natural
parameter, since the solution size is the number of vertices in the input
graph. Instead, we may look at structural parameterizations of
the input graph, for instance its treewidth or clique-width.

A lot of NP-hard graph problems become fixed parameter tractable (FPT, solvable
in $f(k)n^{\bigoh(1)}$-time for parameter-value $k$) when parameterized by
treewidth. Many examples of problems that can be checked locally, e.g.,
Independent Set, Vertex Cover, Dominating Set and so on, are even EPT when
parameterized by treewidth, meaning that the problems can be solved in time
$2^{\bigoh(k)}n^{\bigoh(1)}$~\cite{complexity} (also referred to as having a single exponential algorithm). When parameterized by
clique-width, hardly any of these problems are known to be EPT. For instance
Dominating Set has recently been shown solvable in time $2^{\bigoh(k \log k)}
n^{\bigoh(1)}$ for clique-width $k$~\cite{OSV}, but this is still not EPT. 

\begin{figure}[h]
  \centering
  \includegraphics[scale=0.85]{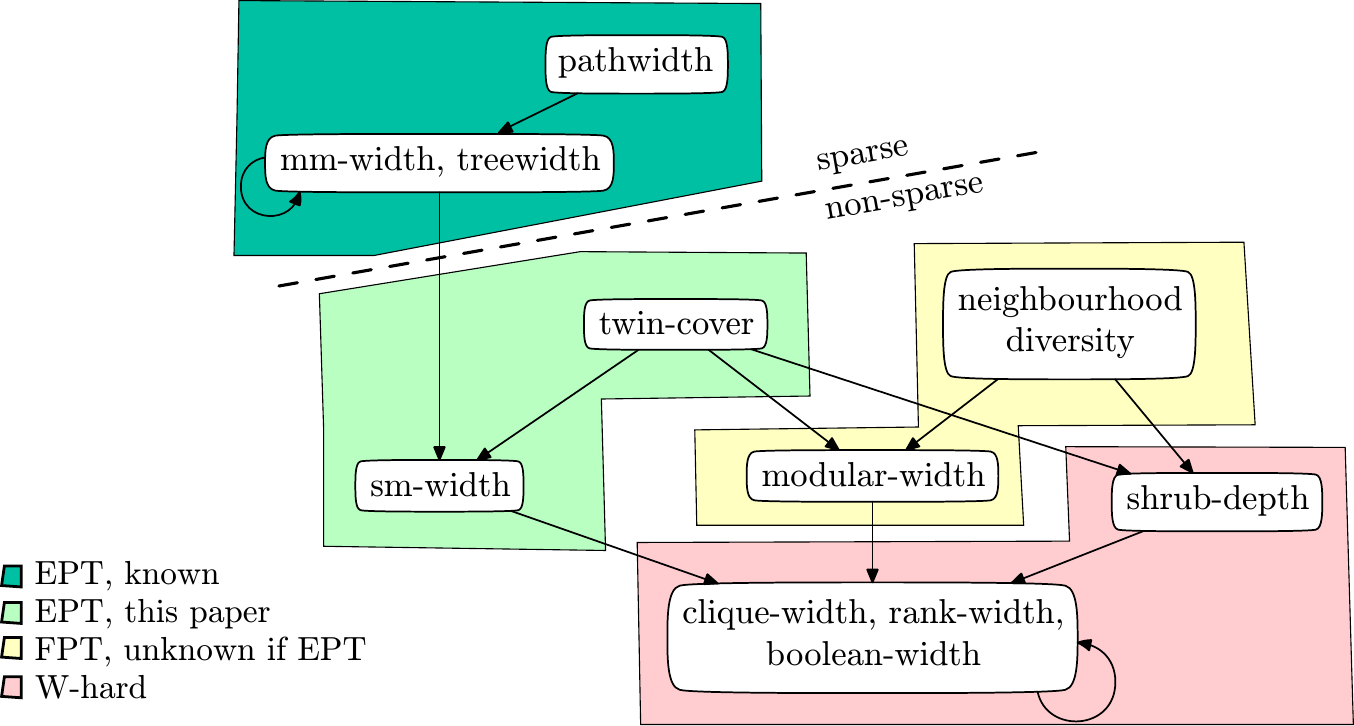}
  \caption{A small overview of how certain graph width parameters
    relate to each other. A directed path from parameter $a$ to $b$
    means that there exists a function $f$ so that for all graphs $G$,
    $b(G) \leq f(a(G))$. The arrows related to $\sm$-width are from
    \cite{ST}, for the others see Gajarsk\'{y} et
    al.~\cite{gajarsky}. The colors depict the complexity of
    Hamiltonian Cycle parameterized with the respective parameters.}
\label{fig:relatingParameters}
\end{figure}

For problems that have a global constraint, like Steiner Tree, Hamiltonian
Cycle and Feedback Vertex Set, EPT algorithms parameterized by treewidth were
for a long time not known. For example, the asymptotically best algorithm for
Hamiltonian Cycle was for a long time the folklore $n^{\bigoh(1)}k^{\bigoh(k)}$
time algorithm, resulting in a belief that graph problems with a global
requirement may not have EPT algorithms.  Recently, however, a breakthrough
paper by Cygan et al.~\cite{cutAndCount} gave a randomized EPT algorithm for
Hamiltonian Cycle, and other problems with global constraints, when
parameterized by treewidth.  Shortly after this, Bodlaender et
al.~\cite{jesper} and then Fomin et al.~\cite{matroidReps}, also found
deterministic EPT algorithms for Hamiltonian Cycle parameterized by treewidth.
Both the papers~\cite{jesper,matroidReps} are general, in the sense that they
provide a framework for solving many problems.  Graph classes of bounded
treewidth are all sparse, so one may wonder if using either of these new
frameworks will help in finding similar EPT results for globally constrained
problems like Hamiltonian Cycle for a parameter bounded also on non-sparse
graph classes. The classical structural graph parameter bounded also on
some non-sparse graphs is clique-width.  Unfortunately, it is unlikely that
such a result exists for clique-width, as Hamiltonian Cycle has been shown to
be W-hard when parameterized by clique-width~\cite{W1hardness}. So, we must
focus on a non-sparse parameter which is less general than clique-width.
Examples of some such parameters are modular-width, shrub-depth, neighbourhood
diversity, twin-cover, and the newly introduced split-matching-width (see
Figure~\ref{fig:relatingParameters}).

In the recent paper~\cite{gajarsky} Gajarsk\'{y} et al. give an FPT
algorithm (but not EPT) for Hamiltonian Cycle parameterized by
modular-width, and show W-hardness when parameterized by shrub-depth. 
Split-matching-width is a new parameter introduced by S\ae{}ther and
Telle~\cite{ST} for which Hamiltonian Cycle is FPT~\cite{ST}. Unlike
modular-width, split-matching-width generalizes treewidth, so it is a good
candidate for applying the framework used for treewidth.

In this paper, we will show that using the framework of
\cite{matroidReps} we can solve Hamiltonian Cycle in time
$2^{\bigoh(k)}n^{\bigoh(1)}$ for parameter $k$ being split-matching-width. The
approach will be similar to that of \cite{ST} in the sense that it consists of
two parts; (1) given a graph $G$, finding a
branch decomposition of low split-matching-width, and then (2) solving
Hamiltonian Cycle on $G$ with a runtime depending on the
split-matching-width of the computed branch decomposition. We will in
this paper improve on the results from \cite{ST} by showing the following two
theorems that when combined results in an EPT algorithm for Hamiltonian Cycle
parameterized by split-matching-width.

\begin{theorem}\label{thm:sum:approx}
  Given a graph $G$ of split-matching-width less than $k$, in
  $n^{\bigoh(1)}2^{\bigoh(k)}$ time we can find a branch decomposition of
  split-matching-width less than $16k$.
\end{theorem}

\begin{theorem}\label{thm:sum:dyn}
  Given a graph $G$ and a branch decomposition of split-matching-width
  $k$, we can decide if $G$ has a Hamiltonian Cycle in time
  $n^{\bigoh(1)}2^{\bigoh(k)}$.
\end{theorem}

Another result of Theorem~\ref{thm:sum:approx} is that we can improve the
runtime of the algorithms for solving Edge Dominating Set, Chromatic
Number, and Max Cut parameterized by split-matching-width described in
\cite{ST}. In fact, under the Exponential Time Hypothesis the asymptotic
runtimes for Max Cut, Hamiltonian Cycle, and Edge Dominating Set become
optimal~\cite{SETHbounds,lokshtanov2013lower}\footnote{see the Appendix for the
lower bound on Edge Dominating Set}.
(I.e., no $n^{\bigoh(1)}2^{\littleoh(k)}$ algorithm exists.)

This paper is organized as follows: In Section~\ref{sec:prelims}, we
give the necessary definitions and background needed for the rest of
the paper and in Section~\ref{sec:hc} we prove
Theorem~\ref{thm:sum:dyn}. In Section~\ref{sec:approx} we prove
Theorem~\ref{thm:sum:approx}, and then we end this paper in
Section~\ref{sec:conc} where we give a short summary.

The symbol \app\ denotes that the proof can be found in the
Appendix.

\section{Preliminaries and terminology} \label{sec:prelims}

\paragraph{Graph and set preliminaries.} We work on simple undirected
graphs $G = (V,E)$ and denote the set of vertices and set of edges of
a graph $G$ by $V(G)$ and $E(G)$, respectively. We use $n$ to denote
the number of vertices of the graph in question. For an edge between
vertices $u$ and $v$, we simply write $uv$. For a path $P$, by writing
$uPv$ we mean that the endpoints of $P$ are $u$ and $v$. For a graph
$G$ and subset $A \subseteq V(G)$, we denote by $G[A]$ the subgraph of
$G$ \emph{induced} by $A$. That is, the vertex set $V(G[A])$ of $G[A]$
is $A$ and the edge set is $E(G[A]) = \{uv \in E(G) : u,v \in A\}$.
For disjoint sets $A, B \subseteq V(G)$, we denote by $G[A, B]$ the
bipartite subgraph of $G$ induced by the pair $(A, B)$. That is
$V(G[A, B]) = A \cup B$ and $E(G[A, B]) = \{uv \in E(G) : u \in A, v
\in B\}$. For a set of vertices $S \subseteq V(G)$, we denote by
$N_G(S)$ all the vertices in $V(G) \setminus S$ adjacent to $S$. We
omit the subscript $G$ in $N_G(S)$ when it is clear from context. For
a single vertex $v$, we write $N_G(v)$ instead of $N_G(\{v\})$.  To
\emph{contract} an edge $uv$ means to replace the vertices $u$ and $v$
by a new vertex $v_{uv}$ adjacent to exactly the same vertices as $u$
and $v$ combined. For a set $A \subseteq V(G)$, when $V(G)$ is clear
from context, we write $\compl A$ to mean the set $V(G) \setminus A$.
For a graph $G$ and subsets $A,B,C \subseteq V(G)$, we say that $C$
separates $A$ and $B$ if there are no paths from $A \setminus C$ to $B \setminus
C$ in $G[\compl C]$.

\paragraph{Hamiltonian Cycle and Certificates.} A Hamiltonian Cycle in
a graph $G$ is a simple cycle in $G$ of size $|V(G)|$. In this paper
we will solve Hamiltonian Cycle using dynamic programming. The
framework we use will be the same as in \cite{ST}. That is, a
\emph{certificate} is a set of edges forming vertex disjoint paths or
a Hamiltonian cycle, and a witness is a Hamiltonian cycle. We denote
by $\cert(X)$ all the certificates $C$ so that $C \subseteq
E(G[X])$. For two disjoint sets $A$ and $B$ and certificates $s_a \in
\cert(A)$ and $s_b \in \cert(B)$, we define the operation
$\conc(s_a,s_b)$ to be the set of all the certificates $s'$ in
$\cert(A \cup B)$ on the form $s' = s_b \cup s_b \cup E'$ for some set $E'
\subseteq E(G[A, B])$. For sets $S_A \subseteq \cert(A)$ and $S_B
\subseteq \cert(B)$ we denote by $\conc(S_A, S_B)$ the union of all
$\conc(S_a, S_b)$ where $S_a \in S_A$ and $S_b \in S_B$. From these
definitions, we see that it is always the case for disjoint sets $A$
and $B$, $\cert(A \cup B) = \conc(\cert(A) , \cert(B))$.

For two sets $A_1, A_2 \subseteq \cert(A)$, we say that $A_1$
\emph{preserves} $A_2$, denoted $A_1 \better_A A_2$, if for all $B \in
\cert(V \setminus A)$, if $\conc(A_1, \{B\})$ contains a witness, then
$\conc(A_2, \{B\})$ contains a witness. %
We note that if $A_1
\better_A A_2$ then for any $C$ disjoint from $A$, and $X \subseteq
\cert(C)$, we have $\conc(A_1, X) \better_{A \cup C} \conc(A_2, X)$.

\paragraph{Splits and split decompositions.}
A \emph{split} of a connected graph $G$ is a bi-partition $(A, B)$ of the
vertices $V(G)$ where $|A|,|B| \geq 2$ and for all $a \in N(B)$, $N(a)
\cap B = N(A)$. That is, all vertices in $A$ adjacent to $B$ have the
same neighbourhood in $B$. Notice that this property holds if and only
if also for all $b \in N(A)$, $N(b) \cap A = N(B)$.  A bi-partition
$(A, B)$ where either $A$ or $B$ consists of at most one vertex is
said to be a \emph{trivial split}.

A graph $G$ having a split $(A, B)$ can be \emph{decomposed} into
smaller graphs $G_A$ and $G_B$ where $G_A$ is the graph $G$ with all
the vertices of $B$ replaced by a single vertex $v$, called a
\emph{marker}, adjacent to the same vertices in $G_A$ as $B$ is
adjacent to in $G$. $G_B$ is in the same way the graph $G$ where we
replace the vertices $A$ by the marker vertex $v$ so that $N_{G_B}(v)
= N_G(A)$. A graph without a split is called a \emph{prime graph}.
Since all graphs of three or less vertices trivially do not contain
any splits, we say that a prime graph on more than three vertices is a
\emph{non-trivial} prime graph.

\begin{figure}[h!]
  \centering 
  \includegraphics[scale=0.8]{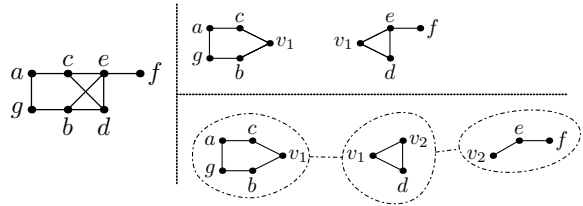}
  \caption{A graph $G$ on vertices $a,b,c,d,e,f,g$ on the left, and a
    the graph $G$ decomposed through the split $(\{a,b,c,g\}.\{d,e,f\})$
    to the upper right. Notice the introduction of marker $v_1$. To
    the lower right we have a split decomposition tree of $G$ with
    markers $v_1$ and $v_2$. In the prime graph with vertices $\{v_1,
    d, v_2\}$ we have $\tot(v_1) = \{a,b,c,g\}$ and $\act(v_1) =
    \{b,c\}$. The weight of $v_2$ in the rightmost prime graph drawn
    in the split decomposition tree is three while in the middle prime
    graph it is one.}
  \label{fig:splitDec}
\end{figure}

A \emph{split decomposition} of a graph $G$ is a recursive
decomposition of $G$ so that all of the obtained graphs are prime.
For a split decomposition of $G$ into $G_1, G_2, \ldots, G_k$, a
\emph{split decomposition tree} is a tree $T$ where each vertex
corresponds to a prime graph and we have an edge between two vertices
if and only if the prime graphs they correspond to share a
marker. That is, the edge set of the tree is $E(T) = \{v_iv_j :
v_i,v_j \in V(T) \text{ and } V(G_i) \cap V(G_j) \neq \emptyset\}$.
See Figure~\ref{fig:splitDec} for an example.
Given a split decomposition of graph $G$ with prime graphs $G_1, G_2,
\ldots, G_k$, we define $\tot(v:G_i)$ recursively to be $\{v\}$ if $v
\in V(G)$, and otherwise to be $\bigcup_{u \in V(G_j) \setminus \{v\}}
\tot(u:G_j)$ for the graph $G_j \not= G_i$ containing the marker $v$
in the split decomposition.  Another way of saying this latter part by
the use of the split decomposition tree $T$ is: if $v$ is not in
$V(G)$, then $\tot(v: G_i)$ is defined to be the vertices of $V(G)$
residing in the prime graphs of the connected component in $T[V(T) \setminus \{
G_i\}]$ where $v$ is also located.  For a set $V' \subseteq V(G_i)$, we
define $\tot(V':G_i)$ to be the union of $\tot(v:G_i)$ for all $v \in
V'$.  We define the \emph{active set} of a vertex $v \in G_i$, denoted
$\act(v:G_i)$ to be the vertices of $\tot(v:G_i)$ that are
contributing to the neighborhood of $v$ in $G_i$. That is,
$\act(v:G_i)$ is defined as $N(V(G) \setminus \tot(v:G_i))$.  Note
that if $G$ has a split decomposition into prime graphs $G_1, \ldots,
G_k$, then for any marker $v$ there are exactly two prime graphs $G_i$
and $G_j$ containing $v$, and we have $\tot(v:G_i) \cup \tot(v:G_j) =
V(G)$. When the prime graph $G_i$ is clear from context, we denote
$\tot(X:G_i)$ and $\act(X:G_i)$ simply as $\tot(X)$ and $\act(X)$.
See Figure~\ref{fig:splitDec} for an example of $\tot()$ and $\act()$.
For a prime graph $G'$ and vertex $v \in V(G')$, when we say the
\emph{weight} of $v$, we mean the cardinality of $\act(v)$.

\paragraph{Split-matching-width.}
A branch decomposition of a set $X$ is a description of recursively
dividing $X$ into smaller and smaller sets until only single element
sets are left. Formally, a branch decomposition of $X$ is a pair $(T,
\delta)$ where $T$ is a subcubic tree (a tree of max degree three) and
$\delta$ is a bijection from the leaves in $T$ to the elements in $X$.
In this paper, when we say a branch decomposition of a graph $G$, we will mean a branch
decomposition of $V(G)$.

A cut of $X$ is a bi-partition of $X$. Each edge $e$ of a branch
decomposition (an edge in the tree) describes a cut/bi-partition of
$X$, namely the cut $(A, B)$ where $A$ are all the elements that are
mapped to from $\delta$ by the leaves on one side of $e$ and $B$ are
the elements mapped by $\delta$ from the leaves of the other side of
$e$. For a set $X$, a \emph{cut function} $f: 2^{X} \rightarrow
\mathbb{N}$ is a symmetric ($f(A) = f(\compl{A})$) function on subsets
of $X$. For a branch decomposition $(T, \delta)$ of $G$ and a cut
function $f$ on $V(G)$, its $f$-width is the maximum of $f(A)$ over
all cuts $(A, \compl{A})$ of $G$ induced by the edges of $T$.  For a
graph $G$, its $f$-width, for a cut function $f$, is the minimum
$f$-width over all branch decompositions of $G$.

The Maximum-Matching-width ($\mm$-width) $\mmw(G)$ of a graph $G$ was
defined by Vatshelle~\cite{vatshelle} based on the cut function $\mm$
defined for $A \subseteq V(G)$ as the size of a maximum matching in
$G[A, V \setminus A]$. 
The split-matching-width ($\sm$-width) $\smw(G)$ of a graph $G$ was
defined by S\ae{}ther and Telle~\cite{ST} based on the cut
function $\sm$ defined using splits and $\mm$ as follows:
\[\sm(A) = 
\begin{cases}
  1  & \text{ if } (A, \compl A) \text{ is a split of } G \\
  \mm(A) &  \text{ otherwise} 
\end{cases}\]%

A function $f:2^X \rightarrow \mathbb{N}$ is said to be
\emph{submodular} if for any subset $A,B \subseteq X$ it holds that $ f(A) + f(B) \geq f(A \cup B) + f(A \cap B)$.

\section{Solving Hamiltonian Cycle in EPT time}\label{sec:hc}

From Theorem~3 of~\cite{matroidReps} applied to a \emph{graphic
  matroid} we have the following corollary, which is the core tool for
getting our EPT algorithm.

\begin{corollary}[\cite{matroidReps}]\label{cor:matroids}
  Let $G$ be a connected graph on $n$ vertices and $S$ a family of
  $p$-sized subsets of $E(G)$. We can for any integer $q$ find a subset
  $\hat S$ of $S$ with $| \hat S | \leq 2^{n}$ so that for any $q$-sized subset
  $Y$ of $E(G)$, if there exists a set $X \in S$ disjoint from $Y$ so
  that $X \cup Y$ is a forest, then there exists a set $\hat X \in
  \hat S$ disjoint from $Y$ so that $\hat X \cup Y$ is a
  forest. Furthermore, the set $\hat S$ can be computed in
  $n^{\bigoh(1)}(|S| 4^n)$ time.
\end{corollary}

This might not seem like something related to finding Hamiltonian
Cycles, since finding a largest forest is polynomial time solvable and
finding a largest cycle is NP-complete. However, as shown in
\cite{matroidReps}, many NP-complete problems that contain a global
connectivity constraint, for instance Steiner Tree and Hamiltonian
Cycle, can be solved faster by the use of
Corollary~\ref{cor:matroids}. In~\cite{matroidReps}, the authors focus
more on the Steiner Tree problem and less on Hamiltonian Cycle, so the
precise usage of Corollary~\ref{cor:matroids} applied to Hamiltonian
Cycle is not very explicit. The following result can, however, be
deduced from their paper, but as it is such a crucial part of our
paper, we need all the details formally stated.

\begin{lemma}[\app{}]\label{lemma:main} Let $G$ be
  a graph and $A \subseteq V(G)$ some set of vertices separated from
  ${V(G)} {\setminus} A $ by $C \subseteq V(G)$ of size at least
  three. Given a family $S$ of subsets of edges of $E(G[A \cup C])$,
  we can in $n^{\bigoh(1)}(|S|12^{|C|})$ time construct a family $\hat{S}
  \subseteq S$ of size at most $6^{|C|}$ so that for any set $Y$ of
  edges in $E(G[C \cup \compl A])$, if there exists a set $X \in S$
  disjoint from $Y$ so that $Y \cup X$ is a Hamiltonian cycle, then
  there exists a set $\hat X \in \hat S$ disjoint from $Y$ so that
  $\hat X \cup Y$ is a Hamiltonian cycle.
\end{lemma}

Lemma~\ref{lemma:main} gives an insight to how it is possible to make EPT
algorithms for Hamiltonian Cycle parameterized by treewidth, as done in
\cite{matroidReps}. The idea is to build a set of partial solutions using
dynamic programming in a bottom up manner in a tree decomposition, and at each
step use Lemma~\ref{lemma:main} to reduce the number of partial solutions needed
to ensure you will find a Hamiltonian cycle in the end of your algorithm. This
works because at each step of the algorithm all partial solutions will be
disjoint paths that have all their endpoints inside a small separator (a bag in
the tree decomposition).  However, when parameterizing by split-matching width,
even for cuts of small $\mm$-value and a vertex cover $C$ of small size, the
partial solutions (certificates) will not necessarily consist of paths that have
endpoints inside $C$, but possibly in $N(C)$, which could be large.  To overcome
this problem, we define what we call an \emph{extension}.

An extension of a certificate is a certificate plus some extra
edges. The idea is that an extension will encompass how a certificate
$C \in \cert(X)$ looks after adding more edges than those in
$E(G[X])$. Formally, for a certificate $C \in \cert(X)$ and set of
edges $E^*$ disjoint from $E(G[X])$, we say that a set $S \subseteq \{
C \cup E' : E' \subseteq E^* \}$ is an extension of $C$ by the set
$E^*$. For a set of certificates $P$, the set $S$ is an extension of
$P$ by $E'$ if it is a union of extensions by $E'$ of the certificates
in $P$.  For a set of certificates $P$ we say that an extension $S$ of
$P$ by $E^*$ is \emph{preserving} if for any edge set $Y$ not
intersecting $E(G[A]) \cup E^*$, if there is a certificate $C \in P$
and $E' \subseteq E^*$ so that $Y \cup C \cup E'$ is a Hamiltonian cycle,
then there is an element $C' \in S$ so that $C' \cup Y$ is a
Hamiltonian cycle. A preserving extension of a single certificate $C$
is simply a preserving extension of $\{C\}$.

\begin{observation} \label{obs:from-ext_p-to-cert_p} For $P \subseteq
  \cert(A)$ and edges $E' \subseteq E(G[A, \compl A])$, if $\mathcal
  S$ is a preserving extension of $P$ by $E'$, then $\{S \setminus E'
  : S \in \mathcal S\}$ preserves $P$.
\end{observation}

Motivated by Observation~\ref{obs:from-ext_p-to-cert_p}, we give the
following lemma, which will be used to reduce the number of
certificates needed to preserve certificate sets over sets of small
$\mm$-value. The result of this is captured in
Corollary~\ref{cor:vc-trimming}. 

\begin{lemma}[\app{}]\label{lemma:pres_ext_onto_C}
  Given a set of certificates $S \subseteq \cert(A)$ and a vertex cover $C$
  of $G[A, \compl A]$ of size at least three, we can in
  $n^{\bigoh(1)}(|S|2^{\bigoh(|C|)})$ time create a preserving
  extension of $S$ by $E^* = E(G[A, C \setminus A])$ of size no
  larger than $6^{|C|}$.
\end{lemma}

\begin{corollary}\label{cor:vc-trimming}
  Given a set $S \subseteq \cert(A)$, and a vertex cover $C$ of $G[A,
  \compl A]$ where $3 \leq |C| = k$, we can in
  $n^{\bigoh(1)}(|S|2^{\bigoh(k)})$ time find a set $\hat S \subseteq
  S$ so that $\hat S \better_A S$ and $| \hat S | \leq 6^k$.
\end{corollary}

Combining Corollary~\ref{cor:vc-trimming} with the result of \cite{ST}
saying that for a split $(A, \compl A)$ and a set $S \subseteq
\cert(A)$ of certificates, we can in time polynomial in
$n^{\bigoh(1)}|S|$-time compute a set $S' \better_A S$ of size
$n^{\bigoh(1)}$, we get the following.

\begin{corollary}\label{cor:trim}
  For a set $A \subseteq V(G)$ and set $S \subseteq \cert(A)$, we can in
  $n^{\bigoh(1)}2^{\bigoh(\sm(A))}$ time compute a set $S' \better_A
  S$ of size $n^{\bigoh(1)}2^{\bigoh(\sm(A))}$.
\end{corollary}

Now that we have defined preserving extensions, and already shown how
we can use this to reduce the size of a preserving set of
certificates, we will show how we can also use extensions to produce
small sets $S$ preserving $\conc(S_1, S_2)$ for certificates $S_1$ and
$S_2$. This is the last step needed to create our dynamic programming
algorithm for Hamiltonian cycle.

\begin{lemma}\label{lemma:join}
  For a tri-partition $(A, B, W)$ of $V(G)$, and $S_a \in \cert(A)$ and
  $S_b \in \cert(B)$, for $k = \max\{\mm(A), \mm(B)\}$ we can in
  $n^{\bigoh(1)}2^{\bigoh(k)}$ time compute a set $S \subseteq \conc(S_a,
  S_b)$ so that $S \better_{A \cup B} \conc(S_a, S_b)$ of size at most
  ${2}^{\bigoh(k)}n^{\bigoh(1)}$.
\end{lemma}

\begin{qedproof}
  The case when both $(A, \compl A)$ and $(B, \compl B)$ are splits,
  we can construct a preserving set $S$ of size polynomial in $n$ in
  $n^{\bigoh(1)}$ time as shown by \cite{ST}. So, we will only give
  proof for the case when at least one of $A$ and $B$ are not
  splits.

  We first assume that there exists a certificate $S_w \in \cert(W)$
  so that the set $\conc(\conc(S_a, S_b), \{S_w\})$ contains a witness
  $H = S_a \cup S_b \cup S_w \cup E'$ where $E'$ is disjoint from the three certificates.  Let $X_a \subseteq A$ and $X_b
  \subseteq B$ be the set of vertices in $A$ and $B$ incident with
  less than two edges of $S_a$ and $S_b$, respectively. That is, $X_a$
  and $X_b$ are exactly the vertices of $A$ and $B$, respectively,
  that are incident with $E'$.
  
  We now show that if a witness $H$ as described above exists, then
  $|X_a| \leq 2\mm(A)$ and $|X_b| \leq 2\mm(B)$. Without loss of
  generality, we prove that this holds for $X_a$. As each vertex in
  $X_a$ must be incident with an edge of $E'$, and each vertex in
  $\compl A$ can be incident to at most two edges of $E'$ since $H$ is
  a simple cycle, there is a matching in $E'$ of at least half the
  size of $X_a$, implying that $|X_a| \leq 2\mm(A)$. The same also
  holds for $X_b$. Let $C$ be a vertex cover of $G[A, \compl A]$. If
  both $\mm(A),\mm(B) \leq k$, then $C \cup X_a \cup X_b$ is a vertex
  cover of size $\bigoh(k)$. This means that by
  Lemma~\ref{lemma:pres_ext_onto_C} that we can construct a preserving
  extension $\hat {S_a}$ of $S_a$ by $E(G[A, (C \cup X_b) \setminus
  A])$ of size $6^{\bigoh(k)}$, which combined with $S_b$ must
  preserve $\conc(S_a, S_b)$. That is, $S' = \{S_a' \cup S_b: S_a' \in
  \hat{S_a}\}$. 

  For the case when either $\mm(A) > \sm(A)$ or $\mm(B) > \sm(B)$, we
  need a slightly different argument. Assume without loss of
  generality that $\mm(B) > \sm(B)$, and thus $(B, \compl B)$ is a
  split. As before $|X_a| \leq 2k$, but now $|X_b|$ is possibly very
  large. What we notice though, is that as each vertex of $X_a$ can be
  incident to at most two edges of $E'$, the number of edges in $E'$ incident
  with $X_a$ is at most $2|X_a|\leq 4k$. This means that no more than $4k$ of the
  paths in $S_b$ will connect directly to $S_b$ by the edges in $E'$.
  As all endpoints of all paths in $S_b$ (including isolated vertices,
  which can be thought of as paths of length zero) have the same
  neighbourhood in $\compl B$ and are interchangeable, we can simply
  disregard all but $4k$ paths of $S_b$, and do the same for the
  remaining $4k$ paths as we did for $S_b$ for the case when there
  were no splits. This means the set $X_b$ is of size at most $8k$
  instead of $4k$, but this constant disappears in the
  $\bigoh$-notation.
\end{qedproof}

We now have the means to prove Theorem~\ref{thm:sum:dyn}. The
following recursive algorithm will, based on Lemma~\ref{lemma:join}
and Corollary~\ref{cor:vc-trimming} decide Hamiltonian cycle in EPT
time given a branch decomposition $(T, \delta)$ of sm-width $k$:

We subdivide an arbitrary edge of $T$ (add a vertex ``in the middle''
of it) and root $T$ in the new vertex $r$ from the subdivision, and
then in a bottom up manner compute for each node $v \in T$ with
children $c_1, c_2 \in T$ a set $S_v \better_{\delta(v)}
\cert(\delta(v))$. We do this by applying Lemma~\ref{lemma:join} on
each pair of certificates of $S_{c_1}$ and $S_{c_2}$ and then bounding
the size of $S_v$ by Corollary~\ref{cor:vc-trimming}. For the base
case where $v \in T$ is a leaf, $\cert(\delta(v))$ can be computed
exactly in polynomial time. In the root node, we have computed $S
\better_{\delta(c_1) \cup \delta(c_2)} \cert(\delta(c_1) \cup
\delta(c_2)) = \cert(V(G))$. Deciding the Hamiltonian cycle problem
can then be answered easily by checking each certificate in $S_{r}$
whether or not it is a Hamiltonian cycle, by polynomial amount of work
for each certificate.  The total amount of work is bounded by the
number of nodes in $T$ (which is $n$), times the work spent at each
node, which by Lemma~\ref{lemma:join} and
Corollary~\ref{cor:vc-trimming} is bounded by
$n^{\bigoh(1)}2^{\bigoh(k)}$.
This concludes the
proof of Theorem~\ref{thm:sum:dyn}.

\section{Approximating sm-width}\label{sec:approx}

In \cite{ST}, S\ae{}ther and Telle gave an algorithm for
constructing an algorithm of split-matching-width $\bigoh(\sm(G)^2)$ in FPT
time. Their procedure consisted of four main steps; 
\begin{enumerate}
\item %
  construct a split decomposition $\mathcal{D}$ of $G$,
\item %
  compute a branch decomposition of $\sm$-width less than $9k$
  for each prime graph in $\mathcal{D}$ where $\smw(G) < k$,
\item %
  adjust each branch decomposition slightly so that the
  lifted-$\sm$-width of each decomposition becomes less than
  $54k^2$, and then finally
\item %
  combine all the branch decompositions together to form a branch
  decomposition for $G$ of $\sm$-width less than $54k^2$.
\end{enumerate}

In this paper we will keep the general structure as in \cite{ST}, but
we will replace steps (2) and (3) by a single step where we compute
branch decompositions for each prime graph of lifted-$\sm$-width
bounded by $18k$ directly. The last part is covered by the following
theorem which can be extracted from the proof of Theorem 13 in~\cite{ST}, 
and the first part was shown by~\cite{cunningham} to be
computable in polynomial time, so we will focus this section on
constructing branch decompositions of low lifted-sm-width for prime
graphs.

\begin{theorem}[\cite{ST}]\label{thm:combiningPrimeDecs}
  Given a graph $G$ and a split decomposition $\mathcal{D}$ over prime
  graphs $G_1, G_2, \ldots$ with corresponding branch decompositions
  $(T_1, \delta_1), (T_2, \delta_2), \ldots$ all of lifted-sm-width
  $k$, we can in polynomial time construct a branch decomposition
  for $G$ of sm-width $k$.
\end{theorem}

The difference between the approach of this paper and of
\cite{ST} is that we in this paper explicitly define the lifted
version of the cut function $\sm$, and show that this cut function
directly can be approximated to a linear factor of $\sm(G)$.

\subsection*{Lifted $\sm$-width}
For a cut function $f$ and prime graph $G'$ of some split
decomposition of $G$, we denote by $f^{\ell}$ the value of $f$
\emph{lifted} from $G'$ to $G$. That is, $f^{\ell} (X) = f(\tot(
{X:G'} ))$. We may also refer to this by simply writing
``lifted-$f$''.
  
\begin{theorem}[\cite{ST}]\label{thm:mm-sub} The cut function
  $\mm$-value is submodular.
\end{theorem}

\begin{theorem}[\cite{sangIlApprox}] \label{thm:oumSeymour} For a
  symmetric submodular cut-function $f$ and graph $G$ of optimal
  $f$-width $k$, a branch decomposition of $f$-width at most $3k+1$
  can be found in $n^{\bigoh(1)}(2^{3k+1})$ time.
\end{theorem}

From the definition of submodularity, and Theorem~\ref{thm:mm-sub}
saying that $\mm$ is submodular, we can deduce that also lifted-$\mm$
is submodular. We simply substitute $\mm^{\ell}(X)$ by $\mm(\tot(X))$
in the submodularity inequality:
  \begin{align*}
    \mm^{\ell}(A) + \mm^{\ell}(B) 
      & = \mm(\tot(A)) + \mm(\tot(B))\\
      & \geq \mm(\tot(A \cup B)) + \mm(\tot(A \cap B))\\ 
      & = \mm^{\ell}(A \cup B) + \mm^{\ell}(A \cap B).
  \end{align*}
\begin{corollary}\label{cor:lifted-mm-sub}
  Lifted-$\mm$-value is submodular.
\end{corollary}

\begin{corollary}[\app{}] \label{cor:lifted-mm-approx}
  For a graph of lifted-$\mm$-width $k$, we can find a branch
  decomposition of $\mm^{\ell}$-width at most $3k+1$ in
  $n^{\bigoh(1)}(2^{3k})$ time.
\end{corollary}

\begin{lemma}[\app{}]\label{aprx:Exists-3k-dec-of-pr.gr}
  Let $G$ be a graph and $\mathcal{D}$ a split decomposition of
  $G$. For any prime graph $G'$ in $\mathcal{D}$ there exists a branch
  decomposition $(T, \delta)$ of $G'$ of $\sm^{\ell}$-width $\leq 3\smw(G)$.
\end{lemma}

The following lemma is an improvement of Lemma~7 of~\cite{ST},
which cascades throughout their paper, improving their analysis to
show that the resulting branch decomposition is of $\sm$-width
$36\sm(G)^2$ instead of $54\sm(G)^2$.  We may note that without the
below lemma, we could use Lemma~7 of~\cite{ST} to get a $27$
approximation instead of a $18$ approximation.

\begin{lemma}[\app{}]\label{lemma:6k:noProof}
  Let $G$ be a graph of split-matching-width less than $k$, and $G'$ a
  non-trivial prime graph in a split decomposition of $G$.  For any
  vertex $v$ in $G'$ of weight at least $3k$, $v$ is either adjacent
  to exactly one other vertex of weight at least $3k$ or the $\mm$-value
  of $\tot(v)$ is less than $6k$.
\end{lemma}

As we notice from Lemma~\ref{lemma:6k:noProof}, vertices of weight
$3k$ are more restricted than the rest of the vertices. We say that
a vertex of weight at least $3k$ is \emph{heavy}, and an edge incident
with two heavy vertices is called a \emph{heavy} edge.
 
For a heavy edge $uv$, if a branch decomposition of lifted-sm-width
less than $3k$ induce a non-trivial cut $(A, B)$, it must be the case
that $u$ and $v$ are either both in $A$ or both in $B$. Otherwise, the
lifted-$\sm$-width will be too large. This means that in any branch
decomposition of lifted-sm-width less than $3k$, $(\{uv\},
\compl{\{uv\}})$ must be a cut induced by the decomposition.

\begin{corollary}\label{aprx:coroll:exists-6k-mm-width}
  By contracting each heavy edge $uv$ in a prime graph $G'$ of $G$ to
  single vertices $v_{uv}$, letting $\tot(v_{uv}) = \tot(\{u,v\})$, we
  get a graph of $\mm^{\ell}$-width at most $6\sm(G)$.
\end{corollary}

Based on Corollary~\ref{aprx:coroll:exists-6k-mm-width}, and
Corollary~\ref{cor:lifted-mm-approx} saying that we can
$3$-approximate the width, we get the a branch decomposition of
$\sm^{\ell}$-width $18\sm(G)$ for each prime graph of $G$.

\begin{lemma}\label{lemma:finalPrimeDec}
  For a graph $G$ and prime graph $G'$ in a branch decomposition of
  $G$, a branch decomposition $(T, \delta)$ over $G'$ of
  lifted-$\sm$-width at most $18$ times the optimal sm-width $k$ of $G$
  can be generated in $n^{\bigoh(1)}(2^{18k})$ time.
\end{lemma}

\begin{qedproof}
  By first contracting each heavy edge as described in
  Corollary~\ref{aprx:coroll:exists-6k-mm-width}, we have a graph of
  lifted-mm-width less than $6k$. By
  Corollary~\ref{cor:lifted-mm-approx}, we can construct a branch
  decomposition of $sm^{\ell}$-width less than $18k$. For each
  contracted heavy edge $uv$, we append $u$ and $v$ to the leaf
  $v_{uv}$ corresponding to the contracted edge. This will only alter
  the decomposition in the form of adding trivial cuts, i.e.,
  splits. And thus, the lifted $\sm^{\ell}$-width of the decomposition
  remains the same.
\end{qedproof}

Lemma~\ref{lemma:finalPrimeDec} completes the part of finding branch
decompositions of the prime graphs with lifted-sm-width only a linear
factor larger than the original graph. Putting
Lemma~\ref{lemma:finalPrimeDec} together with the fact that we can
find a split decomposition in polynomial time by~\cite{cunningham} and
Theorem~\ref{thm:combiningPrimeDecs} saying that we can combine
lifted-sm-decompositions of prime graphs together to form a branch
decomposition of the original graph, we have proved
Theorem~\ref{thm:sum:approx} as promised.

\section{Conclusions}\label{sec:conc}

We have shown a dynamic programming algorithm solving Hamiltonian
Cycle in $n^{\bigoh(1)}2^{\bigoh(k)}$ time when given a branch
decomposition of sm-width $k$. We have also supplied an algorithm for
finding a branch decomposition of a graph $G$ of $\sm$-width
$\bigoh(\sm(G))$ by focusing on lifted-$\sm$-width of prime
graphs. This results in an EPT algorithm for Hamiltonian Cycle. In
fact, combining the algorithm for finding branch decompositions of low
split-matching-width with the three algorithms for solving Chromatic
Number, Edge Dominating Set, and Max Cut given in \cite{ST}, we end up
with algorithms of runtime, $n^{\bigoh(1)}k^{\bigoh(k)}$,
$n^{\bigoh(1)}2^{\bigoh(k)}$, and $n^{\bigoh(1)}2^{\bigoh(k)}$,
respectively, which under the Exponential Time Hypothesis is optimal\cite{SETHbounds,lokshtanov2013lower}\footnote{For Edge Dominating Set lower bound, see the Appendix.} (no algorithm where the $\bigoh(k)$'s are exchanged with $\littleoh(k)$'s exist).

\bibliography{ref}

\begin{thebibliography}{10}

\bibitem{jesper}
H.~L. Bodlaender, M.~Cygan, S.~Kratsch, and J.~Nederlof.
\newblock Deterministic single exponential time algorithms for connectivity
  problems parameterized by treewidth.
\newblock In {\em Proceedings ICALP}, pages 196--207. Springer, 2013.

\bibitem{cunningham}
W.~H. Cunningham.
\newblock Decomposition of directed graphs.
\newblock {\em SIAM Journal on Algebraic Discrete Methods}, 3(2):214--228,
  1982.

\bibitem{cutAndCount}
M.~Cygan, J.~Nederlof, M.~Pilipczuk, J.~M.~M. van Rooij, and J.~O. Wojtaszczyk.
\newblock Solving connectivity problems parameterized by treewidth in single
  exponential time.
\newblock In {\em Proceedings FOCS}, pages 150--159. IEEE, 2011.

\bibitem{complexity}
J.~Flum and M.~Grohe.
\newblock {\em Parameterized complexity theory}, volume~3.
\newblock Springer, 2006.

\bibitem{W1hardness}
F.~V. Fomin, P.~Golovach, D.~Lokshtanov, and S.~Saurabh.
\newblock Intractability of clique-width parameterizations.
\newblock {\em SIAM Journal on Computing}, 39(5):1941--1956, 2010.

\bibitem{matroidReps}
F.~V. Fomin, D.~Lokshtanov, and S.~Saurabh.
\newblock Efficient computation of representative sets with applications in
  parameterized and exact algorithms.
\newblock In {\em Proceedings SODA}, pages 142--151, 2014.

\bibitem{gajarsky}
J.~Gajarsk{\`y}, M.~Lampis, and S.~Ordyniak.
\newblock Parameterized algorithms for modular-width.
\newblock In {\em Proceedings IPEC}, pages 163--176. Springer, 2013.

\bibitem{SETHbounds}
D.~Lokshtanov, D.~Marx, and S.~Saurabh.
\newblock Known algorithms on graphs of bounded treewidth are probably optimal.
\newblock In {\em Proceedings SODA}, pages 777--789. SIAM, 2011.

\bibitem{lokshtanov2013lower}
D.~Lokshtanov, D.~Marx, S.~Saurabh, et~al.
\newblock Lower bounds based on the exponential time hypothesis.
\newblock {\em Bulletin of EATCS}, 3(105), 2013.

\bibitem{gareyJohnson}
R.~G. Michael and S.~J. David.
\newblock Computers and intractability: a guide to the theory of
  np-completeness.
\newblock {\em WH Freeman \& Co., San Francisco}, 1979.

\bibitem{OSV}
S.~Oum, S.~H. S{\ae}ther, and M.~Vatshelle.
\newblock Faster algorithms for vertex partitioning problems parameterized by
  clique-width.
\newblock {\em Theoretical Computer Science}, 535:16--24, 2014.

\bibitem{sangIlApprox}
S.~Oum and P.~Seymour.
\newblock Approximating clique-width and branch-width.
\newblock {\em Journal of Combinatorial Theory, Series B}, 96(4):514--528,
  2006.

\bibitem{ST}
S.~H. S{\ae}ther and J.~A. Telle.
\newblock Between treewidth and clique-width.
\newblock {\em to appear in proceedings of WG 2014}, 2014.
\newblock Invited to contribute to special section of Algorithmica.

\bibitem{restricted3SAT}
C.~A. Tovey.
\newblock A simplified np-complete satisfiability problem.
\newblock {\em Discrete Applied Mathematics}, 8(1):85--89, 1984.

\bibitem{vatshelle}
M.~Vatshelle.
\newblock {\em New width parameters of graphs}.
\newblock PhD thesis, The University of Bergen, 2012.

\bibitem{EDSreduction}
M.~Yannakakis and F.~Gavril.
\newblock Edge dominating sets in graphs.
\newblock {\em SIAM Journal on Applied Mathematics}, 38(3):364--372, 1980.

\end{thebibliography}
\bibliographystyle{plain}

\clearpage
\appendix
\section{Appendix}
\begin{proposition}
  If for some $\delta_0 > 0$ there does not exist a $2^{(\delta_0 n)}
  n^{\bigoh(n)}$ algorithm for 3-SAT, then there exists a $\delta_1 > 0$ so that
  no $2^{(\delta_1 n)} n^{\bigoh(n)}$ algorithm exists for Edge Dominating Set. 
\end{proposition}
 
\begin{qedproof}
This follows from looking at the NP-hardness reduction from 3-SAT to 3-SAT where
each variable occurs at most 3 times~\cite{restricted3SAT}, and the NP-hardness reduction from 3-SAT
where each variable occurs at most 3 times to Edge Dominating Set
in bipartite graphs of maximum degree 3~\cite{EDSreduction}. The resulting Edge
Dominating Set instance $(G,k)$ resulting of applying these two reductions on
3-SAT formula $\phi$ has its number of
vertices bounded by a constant factor $c$ times the number of variables in $\phi$. 
So, any $2^{\delta_1 n}n^{\bigoh(1)}$ algorithm for Edge Dominating Set implies
a $2^{\frac{\delta_1}{c}n}n^{\bigoh(1)}$ algorithm for 3-SAT. 
\end{qedproof}

\begin{lemma} \label{lemma:2^n_hc} Given a graph $G$ of $k$ vertices,
  and family $S$ of edge-sets, we can in $n^{\bigoh(1)}(|S|{12}^{k})$ time
  find a subset $\hat{S}$ of $S$ of size at most $6^k$ so that for any
  set $Y$ of edges in $E(G)$, if there exists a set $s \in S$ disjoint
  from $Y$ so that $s \cup Y$ form a Hamiltonian cycle, then there is a
  set $\hat{s} \in \hat{S}$ disjoint from $Y$ so that $s \cup Y$ form a
  Hamiltonian cycle.
\end{lemma}

\begin{qedproof}
  For a set $x \in S$ let $D_i(x)$ denote the set of vertices in
  $V(G)$ incident to exactly $i$ of the edges in $x$, i.e., the
  vertices of degree $i$ in the graph $(V(G), x)$). We notice that if
  $Y \cup s$ is a Hamiltonian cycle, then $s$ consists only of vertex
  disjoint paths. Furthermore, $Y \cup s'$ is a Hamiltonian cycle only if
  $D_i(s) = D_i(s')$ for $i = 0,1,2$. So, to produce a smaller subset
  $\hat S$ of $S$ as stated in the lemma, we start by categorizing
  each of the $s \in S$ that consist of only vertex disjoint paths
  into one of at most $3^{|V(G)|}$ classes $S[D_0][D_1][D_2]$,
  depending on the content of $D_i(s)$ (we put $s$ into the class
  where $D_0 = D_0(s), D_1 = D_1(s)$ and $D_2 = D_2(s)$). One result
  of this is that all sets in the same class will consist of exactly
  the same number of edges.

  Now we apply Corollary~\ref{cor:matroids} on each of the classes
  $S[D_0][D_1][D_2]$ with $p$ being the size of the edge sets in the
  particular class, and $q = n-p-1$. We put the result into $\hat
  S[D_0][D_1][D_2]$ for each tri-partition $D_0, D_1, D_2$. We will
  now show that when $\hat S$ is the union of each
  ${\hat{S}[D_0][D_1][D_2]}$, then $\hat S$ satisfies the statement.

  First of all, the size of $\hat S$ is at most $3^k2^k$, since there
  are at most $3^k$ equivalence classes, and each class contributes by
  at most $2^k$ sets.

  Second, we will have to show for every $Y \subseteq E(G)$ so that
  there exists a set $s \in S$ disjoint from $Y$ where $Y \cup s$ is a
  Hamiltonian cycle of $G$, there also exists a set $\hat s \in \hat
  S$ so that also $Y \cup \hat s$ is a Hamiltonian cycle:

  Suppose for disjoint $s$ and $Y$, $Y \cup s$ is a Hamiltonian cycle of
  $G$. This means $D_0(s) = D_2(Y)$, $D_1(s) = D_1(Y)$ and $D_2(s) =
  D_0(Y)$, and $s$ is an element of $S[D_0(s)][D_1(s)][D_2(s)]$. We
  claim that for all other $s' \in S[D_0(s)][D_1(s)][D_2(s)]$ the set
  $Y \cup s'$ is a cycle cover of $G$. That is, each vertex in $V(G)$ is
  adjacent to exactly two edges in $Y \cup s'$. We will show that there
  exists a $s' \in \hat S[D_0(s)][D_1(s)][D_2(s)]$ so that the cycle
  cover $Y \cup s'$ is connected, and hence is a Hamiltonian cycle. 
  Let $e$ be anyi edge
  in $Y$, and let $Y_e$ be $Y \setminus \{e\}$. We notice that $s \cup Y_e$ is a
  (Hamiltonian) path and hence a subtree in $G$. That means by
  Corollary~\ref{cor:matroids}, there must be a $\hat s \in \hat
  S[D_0(s)][D_1(s)][D_2(s)]$ so that $\hat s \cup Y_e$ is a forest. As we
  know all the elements of $S[D_0(s)][D_1(s)][D_2(s)]$ have the same
  number of edges, $H' = \hat s \cup Y_e$ must contain $|V(G)| - 1$
  edges. This means $H'$ is one single component, as we know $H'$ is
  acyclic. Furthermore, we know $H' \cup \{e\} = \hat s \cup (Y_e \cup \{e\})$ is a
  cycle cover, so $H' \cup \{e\} = \hat s \cup Y$ is indeed a Hamiltonian cycle
  of $G$.
\end{qedproof}

To prove Lemma~\ref{lemma:main}, we use what is called a
\emph{torso}. For a graph $G$ and subset $S \subseteq V(G)$, we say
that the \emph{torso} of $G$ in $S$ is the graph we get by taking
$G[S]$ and then add the edges $\{uv : \exists uPv \in G \text{ s.t. }
P \in G \setminus S\}$. In general this means we take $G[S]$ and for
each component $C \in G \setminus S$ make $N(C)$ into a clique. In
this we speak of a torso in the following proof, the components $C \in
G \setminus S$ will only have two neighbours in $C$, and hence each
component will only contribute to the torso by a single edge. %

\begin{figure}[h!]
  \centering
    \begin{minipage}[h]{.45\linewidth}
      \includegraphics[page=1,width=\textwidth]{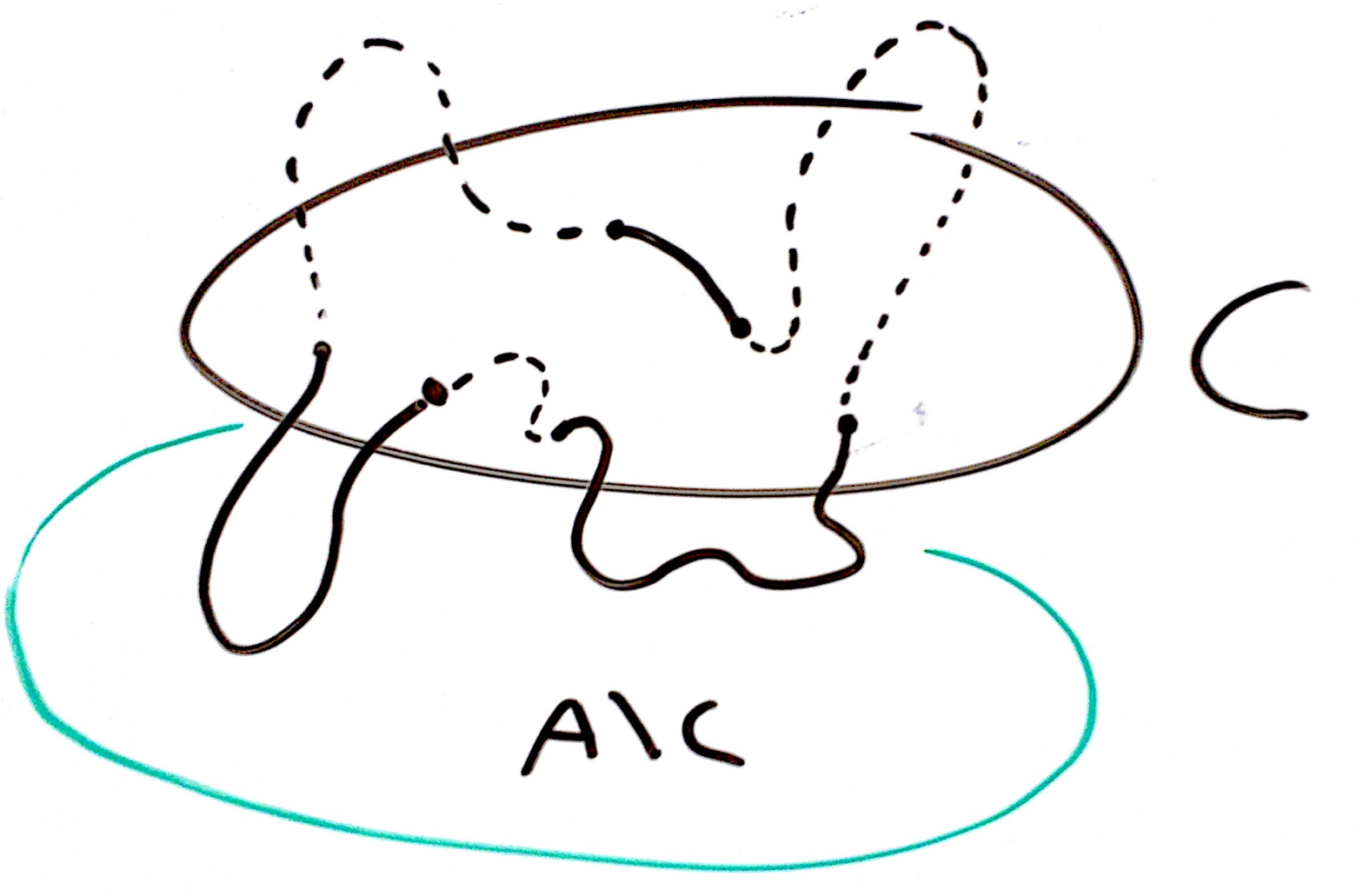}
    \end{minipage} $\Rightarrow$
    \begin{minipage}[h]{.45\linewidth}
      \includegraphics[page=2,width=\textwidth]{torso2.pdf}
    \end{minipage}
    \caption{As described in the proof of
      Lemma~\ref{lemma:main}. The dotted paths are $Y$ and the solid
      paths constitute $X$. We see that if $Y \cup X$ is a Hamiltonian
      cycle in $G$ (the left), then the torso of $Y$ plus the torso of
      $X$ in $C$ must form a Hamiltonian cycle in the torso of $G$.}
    \label{fig:torso}
\end{figure}

\begin{qedproof}[Proof of Lemma~\ref{lemma:main}.]

  Let $G'$ be the complete graph on $V(G)$ (i.e., $V(G') = V(G)$ and
  $E(G') = \{uv : u \in V(G), v \in V(G)\}$).  We notice that for
  disjoint sets $X \subseteq E(G[A])$ and $Y \subseteq E(G[\compl A
  \cup C]) \,$ $X \cup Y$ form a Hamiltonian cycle in $G$ only if $X
  \cup Y$ also form a Hamiltonian cycle in $G'$. Also, $X \cup Y$ can
  be a Hamiltonian cycle only if for some $Y' \subseteq E(G'[C])$ $X
  \cup Y'$ is a Hamiltonian cycle (namely the torso $Y'$ of $Y$ in
  $C$). So, we will from here on assume $V(G') = V(G) = A \cup C$.

  Our goal now will be to reduce the problem from looking at $G'[A
  \cup C]$ to only looking at $G'[C]$. First notice that unless $X$
  induces a disjoint set of paths ending in $C$ which covers all the
  vertices in $A \setminus C$, then $X \cup Y$ cannot be a Hamiltonian
  cycle for any $Y \subseteq E(G'[C])$. Let $T \subseteq S$ be the set
  consisting of all $X \in S$ where $X$ is consisting of such edge
  disjoint set of paths. For any $X \in T$ and $Y \subseteq
  E(G'[C])\,$ $X \cup Y$ is a Hamiltonian cycle of $G'[A \cup C]$ if
  and only if for the torso $X'$ of $X$ in $X$, $X' \cup Y$ is a
  Hamiltonian cycle of $G'[C]$, as depicted in
  Figure~\ref{fig:torso}. Therefore, to construct the set $\hat S$ in
  the statement of the lemma, we do the following:
  
  We first let $T'$ be the set of these torso's. By first removing all
  the duplicates in $T'$ and then applying Lemma~\ref{lemma:2^n_hc} to
  it, we get a subset $\hat {T'}$ of size at most $6^{|C|}$. The set
  $\hat S = \{X \in T: X' \in \hat {T'} \}$ will thus be a set of size
  at most $6^{|C|}$ and such that for any $X \in S$ and $Y \subseteq
  E(G[\compl A \cup C])$, if $X$ and $Y$ are disjoint and $X \cup Y$
  is a Hamiltonian cycle of $G$, then $\exists \hat X \in \hat S$
  disjoint from $Y$ so that $\hat X \cup Y$ is a Hamiltonian cycle in
  $G$.
\end{qedproof}

\begin{qedproof}[Proof of Lemma~\ref{lemma:pres_ext_onto_C}]
  We prove this first for a single certificate $S = \{G'\}$ and then
  generalize to a set $S$ containing multiple certificates in the end
  of the proof.

  Suppose $Y$ is a set of edge disjoint paths so that $Y \cup G' \cup E'$ is
  a Hamiltonian cycle for some $E' \subseteq E^*$. We know neither
  $E'$ nor $Y$ consist of edges in $E(G[A])$. So, by temporarily
  removing the edges of $E(G[A])$ not in $E(G')$ we get a new graph
  $G^*$ for which $Y \cup G' \cup E'$ is also a Hamiltonian cycle. Clearly,
  all Hamiltonian cycles of $G^*$ are also Hamiltonian cycles of $G$.
  As was observed in~\cite{ST}, as all paths in $G'$ must be incident
  with an edge of $E'$ and each edge in $E'$ must be incident with a
  vertex in $C$, yet each vertex in $C$ can be incident with at most
  two edges of $E'$, there cannot be more than $2|C|$ paths in
  $G'$. Now let $X$ be the endpoints of the paths in $G'$.  Each path
  has at most $2$ endpoints, so $|X|$ is at most $4|C|$. The set $X
  \cup C$ is of size at most $5k$ and disconnects $G'$ from the rest
  of the graph. We then do the following subroutine to construct a
  preserving extension. 
  
  \begin{tabbing}
    xx\=xx\=xx\=xx\= \kill
    \> $S' = \{G'\}$\\
    \> for each $e \in E'$ incident with $X$:\\
    \>\> for each $s \in S'$ add $s \cup \{e\}$ to $S'$\\
    \>\> reduce the size of $S'$ using Lemma~\ref{lemma:main} with $X
    \cup C$ as separator\\
    \> end for
  \end{tabbing}
  The set $S'$ resulted from this subroutine will be a preserving
  extension of $G'$ by $E'$, and by Lemma~\ref{lemma:main}, its size
  will not exceed $2^{|C|+|X|} \leq 2^{5|C|}$. The runtime to create
  this set will be $n^{\bigoh(1)}(2^{\bigoh(|C|)})$.

  For a set $S$ of multiple certificates, we find for each $G' \in S$
  a preserving extension $S_{G'}$ by $E'$ and then reduce the size of
  their union to $6^{|C|}$ by Lemma~\ref{lemma:main}, now using only
  $C$ as the separator. The total runtime will now be
  $n^{\bigoh(1)}(|S|2^{\bigoh(|C|)}$ still since it will take
  $n^{\bigoh(1)}(|S|2^{\bigoh(|C|)})$ time to create the union and
  $n^{\bigoh(1)}(|S|2^{\bigoh(|C|)})$ time to reduce the size of this
  union.
\end{qedproof}

\begin{qedproof}[Proof of Lemma~\ref{lemma:6k:noProof}]
  It has already been shown in~\cite{ST} that if $v$ has weight
  $3k$ it cannot be adjacent to two other vertices of weight $3k$. So,
  we only need to prove that if $v$ is not adjacent to any other
  vertex of weight $3k$ or more, the $\mm$-value of $\tot(v)$ is less
  than $6k$.

  We know from Lemma~\ref{aprx:Exists-3k-dec-of-pr.gr} that there must
  exist a branch decomposition of $G'$ of $\sm^{\ell}$-width less than
  $3k$, and all non trivial cuts have $\mm^{\ell}$-value less than
  $3k$. This means there must be a tri-partition $(\{v\}, A, B)$ of
  $V(G')$ so that $\sm^{\ell}(\{v\})$, $\sm^{\ell}(A)$, and
  $\sm^{\ell}(B)$ are all less than $3k$.
  
  Without loss of generality, we show that the number of vertices in
  $\tot(A)$ adjacent to $\tot(v)$ is less than $3k$. Since $(A, \{v\}
  \cup B)$ cannot be a split, $A$ must either consist of a single
  vertex, or have $\sm^{\ell}(A) = \mm^{\ell}(A)$. If $A$ consist of a
  single vertex, by our assumptions that $v$ was not adjacent to any
  vertex of weight $3k$ or more, we have $|N(\tot(v)) \cap \tot(A)| <
  3k$. If $A$ consists of more than a single vertex, we know
  $\mm^{\ell}(A) = \sm^{\ell} < 3k$. Since any minimum vertex cover of
  $G\left[\tot(A),\, \tot(\{v\} \cup B)\right]$ must either consist of
  all of $\act(v)$ or none of it (since $\act(v)$ are twins in this
  bipartite graph), and $|\act(v)| \geq 3k > \mm^{\ell}(A)$, the
  minimum vertex cover of $G\left[\tot(A),\, \tot(\{v\} \cup
    B)\right]$ must consist of all the vertices of $A$ adjacent to
  $\tot(v)$. Hence, $|N(\tot(v)) \cap \tot(A)| \leq \mm^{\ell}(A) \leq
  3k$. The same arguments applies to $B$, so we can conclude that
  \begin{align*}
    |N(\tot(v))| &\leq  |N(\tot(v)) \cap A| + |N(\tot(v)) \cap B| \\
    &< 3k + 3k \enspace.
  \end{align*}
  And so the $\mm$-value of $\tot(v)$ is no more than $6k$ as
  $N(\tot(v))$ is a vertex cover of $(\tot(v), \compl{\tot(v)})$.
\end{qedproof}

\begin{figure}[h!]
  \centering
  \includegraphics[]{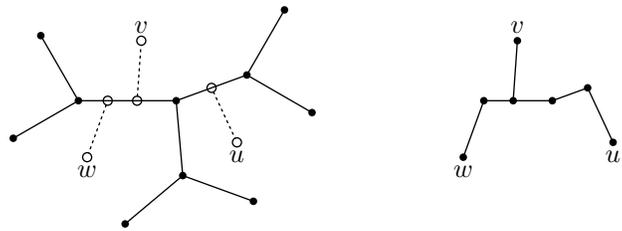}
  \caption{As described in the proof of
    Lemma~\ref{aprx:Exists-3k-dec-of-pr.gr}. $u$,$v$, and $w$ are
    vertices of the prime graph and the dotted lines are the pendants
    added to $T$ (as described in the proof). To the right is the
    resulting decomposition $(T',\delta')$}
  \label{fig:lifted}
\end{figure}

\begin{qedproof}[Proof of Lemma~\ref{aprx:Exists-3k-dec-of-pr.gr}]
  We will give a proof by construction. Let $(T', \delta')$ be a
  branch decomposition of $G$ of optimal split-matching width. We will
  transform $(T, \delta)$ into a branch decomposition $(T', \delta')$
  for $V(G')$. For each vertex $v \in V(G')$ of weight one, we simply
  replace the mapping to $\act(v)$ by a mapping to $v$. For the
  vertices $v \in V(G')$ of weight two or more, there will always be
  an edge $e_v$ in $T$ where at least one third of $\act(v)$ is on
  each side of the cut induced by $e_v$. For each vertex $v \in V(G')$
  of weight at least two, we append a vertex mapping to $v$ to such an
  edge $e_v$ as depicted in Figure~\ref{fig:lifted}. We then
  repeatedly remove all leaves not mapping to a vertex in $V(G')$, so
  that we are left with a branch decomposition $(T', \delta')$ of
  $V(G')$ (see Figure~\ref{fig:lifted}). We now claim that this branch
  decomposition has lifted-$\sm$-width no more than $3\sm(G)$.

  For each trivial cut $(\{v\}, V(G') \setminus \{ v\})$ of
  $(T',\delta')$ the lifted-$\sm$-value is $1$, as it is either a
  split or trivial-split.  For each non-trivial cut $(A', B')$ of
  $(T', \delta')$, there is an associated non-trivial cut $(A, B)$ in
  $(T, \delta)$ where for each $v \in A'$/$B'$ at least one third of
  $\act(v)$ is in $A$/$B$. Let $C$ be a minimum vertex cover of
  $G[A,B]$. For a minimum vertex cover, two twins are either both in
  the cover, or both not in the cover, so for each vertex $v \in
  A$/$B$ either one third of $\act(v)$ is in $C$ or none of $\act(v)
  \cap A$/$B$ is in $C$. Let $C'$ be the vertices $v \in V(G')$ for
  which at least one third of $\act(v)$ is in $C'$. The size of
  $\act(C')$ is at most three times $C$ and we will now show that
  $\act(C')$ in fact is a vertex cover of $G[\tot(A), \tot(B)]$,
  proving that the lifted-$\sm$-width of $G'$ is at most $3\sm(G)$.

  There is an edge between $a$ and $b$ in $G[A',B']$ if and only if
  there is an edge between $\tot(a)$ and $\tot(b)$ in $G[\tot(A),
  \tot(B)]$. So, if for all edges $uv$ in $G[A', B']$, either $u$ or
  $v$ is in $C'$, we are done. Assume $uv \in E(G[A',B']$ where $u \in
  A'$ and $v \in B'$. This means one third of $\act(u)$ and one third
  of $\act(v)$ is on the opposite side of each other in $(A,
  B)$. However, as $\act(u)$ are twins and adjacent to all of
  $\act(v)$, this means one third of either $\act(u)$ or $\act(v)$
  must be in $C$, and hence $u$ or $v$ must be in $C'$.
\end{qedproof}

\end{document}